\documentclass[amsmath, a4paper, prl, twocolumn, showpacs]{revtex4}

\usepackage{amssymb}
\usepackage{amsmath}
\usepackage{color}
\usepackage{graphics, graphicx}
\usepackage{bbold}
\usepackage{psfrag}
\usepackage{mathcomp}

\begin{document}
\title{Effect of interactions on harmonically confined Bose-Fermi mixtures in optical lattices}
\author{Michiel Snoek$^{1}$}
\author{Irakli Titvinidze$^{2}$}
\author{Immanuel Bloch$^{3, 4}$}
\author{Walter Hofstetter$^{2}$}
\affiliation{$^{1}$
Institute for Theoretical Physics, University of Amsterdam, 1090 GL Amsterdam, The Netherlands 
 \\ $^2$Institut f\"ur Theoretische Physik, Johann Wolfgang Goethe-Universit\"at, 60438 Frankfurt/Main, Germany \\ 
$^3$Max-Planck-Institut f\"ur Quantenoptik, Hans-Kopfermann-Str. 1, 85748 Garching, Germany \\
$^4$Ludwig-Maximilians-Universit\"at, Schellingstr. 4/II, 80799 M\"unchen, Germany}


\begin{abstract}
We investigate a Bose-Fermi mixture in a three-dimensional optical lattice, trapped in a harmonic potential. Using Generalized Dynamical Mean-Field theory, which treats the Bose-Bose and Bose-Fermi interaction in a fully non-perturbative way, 
we show that for experimentally relevant parameters a peak in the condensate fraction close to the point of vanishing Bose-Fermi interaction is reproduced within a single band framework. We identify two physical mechanisms contributing to this effect: the spatial redistribution of particles when the interspecies interaction is changed and the reduced phase space for strong interactions, which results in a higher temperature at fixed entropy. 
\end{abstract}

\maketitle

Cold atomic gases offer the exciting possibility to realize mixtures of fermions and bosons in optical lattices \cite{Sengstock2,
Sengstock5,
Guenter,
Best_Bloch}. 
This yields a very interesting system without a clear analog in conventional solid state physics.
The depth of the optical lattice and the Bose-Fermi interaction can be tuned independently, which has made it possible to explore this system experimentally in a detailed way \cite{Best_Bloch}. 
One of the key questions that has been investigated is the effect of the fermions on the mobility of the bosons.
When the fermions are slow compared to the bosons, they act as (dynamical) impurities. Fast fermions, on the other hand, mediate a long-range interaction between the bosons. In both cases  the fermions induce a shift of the bosonic superfluid-Mott 
transition, which has been intensively studied by different experimental groups \cite{Sengstock2, Guenter, Best_Bloch}.
Most importantly, these time-of-flight experiments showed significant loss of bosonic coherence when fermions are present, thus indicating that adding the fermions stabilizes the Mott insulator phase \cite{Sengstock2, Guenter}. A pronounced peak in the bosonic visibility as a function of the Bose-Fermi interaction was observed close to vanishing interaction \cite{Best_Bloch}.
Theoretical studies trying to explain this phenomenon \cite{Albus, Kollath, Demler, Sarma, Varney08, Luhmann08, Lutchyn08, Tewari09, Sengstock3}, however, predict that within a single-band approximation at zero temperature the Mott region shrinks
 when fermions are added. Heating is proposed as an explanation of this apparent contradiction \cite{Sengstock3} and also multiband-effects can lead to an extension of the Mott insulating regime (in a homogeneous system), due to the self-trapping effect \cite{Luhmann08, Lutchyn08, Tewari09, Mering10}. However, in the experiment reported in \cite{Best_Bloch} the increase of visibility happens for small Bose-Fermi interaction, where multi-band effects are supposed to be negligible.

In this Letter we investigate this problem by means of Generalized Dynamical-Mean Field Theory (GDMFT) \cite{Titvinidze08}. This is a non-perturbative method which is well-controlled in three dimensions. We will show that in the homogeneous limit in the single-band framework we indeed reproduce the shrinking of the Mott lobes corresponding to an induced attractive interaction between the bosons mediated by the fermions. To reproduce the experimental results we include two essential physical ingredients.
First of all, we take into account the (experimentally always present) confining potential, 
which leads to an inhomogeneous system. This allows us to probe the interaction-dependent distribution of particles in the trap and its effect on the bosonic coherence. 
Secondly, we account for the fact that the experiments are not performed at constant temperature, but at constant entropy (assuming that the ramp-up of the optical lattice is adiabatic). For strong attractive/repulsive interactions the phase space is strongly reduced, which results in a lower entropy at constant temperature and hence a higher temperature at constant entropy. We will indeed show that the temperature has a minimum close to the position of the peak in the condensate fraction. A combination of these two effects is crucial for a qualitative understanding of the experimental observations.

We study a Bose-Fermi mixture within the single-band approximation and the tight-binding limit, described by the Hamiltonian
\begin{eqnarray}
\label{Hamiltonian}
{\mathcal H}&=&-\sum_{\langle ij \rangle}\left\{ t_{f}c_{i}^{\dagger}c_{j}+t_{b}b_{i}^{\dagger}b_{j} \right\} 
+ \sum_{i, \alpha = b,f} (V_{\alpha i} - \mu_{\alpha}) n_{i}^{\alpha}
\nonumber \\
&& +\sum_{i}\left\{ 
\frac{U_{b}}{2}n^{b}_{i}(n^{b}_{i}-1) + U_{bf}n^{b}_{i}n^{f}_{i}\right\},
\end{eqnarray}
where $c_{i}^{\dagger}$ ($b_{i}^{\dagger}$) is the fermionic (bosonic) creation operator at site $i$, while  
$n^{f}_{i}= c_{i}^{\dagger}c_{i}$  ($n^{b}_{i}= b_{i}^{\dagger}b_{i}$) denotes the corresponding number operator. $\mu_{f(b)}$ is the chemical potential for fermions (bosons). $U_{b}$ and $U_{bf}$ are the on-site boson-boson and boson-fermion interaction parameters, which are related to the bosonic scattering length $a_b$ and the Bose-Fermi scattering length $a_{bf}$ by the overlap integral of the corresponding Wannier functions. The summation index $\langle i,j \rangle$ denotes summation over nearest neighbors, and $t_{f(b)}$ is the tunneling amplitude for fermions (bosons).  

The harmonic trapping potential is incorporated through the term $V_{\alpha i} = V_\alpha | {\bf r}_i - {\bf r}_{\alpha0}|^2$ ($\alpha = b,f$).  
Here we assume the positions of the minima to be equal (${\bf r}_{f0} = {\bf r}_{b0}$), thus ignoring a possible differential sag between the two clouds.
The parameters we have chosen in the simulations for trapped mixtures are closely related to the experimental ones in Ref. \cite{Best_Bloch}. The two species in this experiment are bosonic $^{87}$Rb and fermionic $^{40}$K. By working at an optical lattice wavelength $\lambda = 755$ nm, it is ascertained to have equal dimensionless lattice depth for both species $V_{\rm OL}^b/E_R^b = V_{\rm OL}^f/E_R^f$ ($V_{\rm OL}^\alpha$ being the optical lattice depth and $E_R^\alpha$ the recoil energy). 
This fixes the ratio between the hopping constants as $t_f/t_b =  m_b/m_f = 2.2$. The intrabosonic repulsion is fixed by the bosonic background scattering length of $a_b \approx 100 a_0$ ($a_0$ being the Bohr radius), whereas the interspecies scattering length is tunable by a Feshbach resonance. We fix the total number of particles to approximately $N_b = 4 \times 10^5$ and $N_f = 3\times10^5$, since these numbers yielded the most pronounced experimental signal.
The experimental trap is pancake-shaped.
Since the dipole trap is counter-acted by the laser profile of the blue-detuned lattice beam, the precise trapping frequencies depend on the lattice depth. 
Our simulations are performed in an isotropic three-dimensional trap with trapping frequencies 
equal to the averaged experimental frequencies $\bar \omega_\alpha = \sqrt[3]{(\omega_\perp^\alpha)^2 \omega_z^\alpha}$.

The single band approximation made here cannot be justified for the entire range of interactions realized in the experiments. Significant multi-band effects show up for large attraction between bosons and fermions \cite{Luhmann08, Lutchyn08}. However, we focus on the regime of small Bose-Fermi interactions, where multi-band effects are supposed to be weak. In this way we are able to  isolate 
the effects of the temperature 
and the spatial redistribution of 
the particles and to distinguish them from possible multi-band effects.

To solve this interacting many-body system, we use GDMFT, which is a combination of Gutzwiller mean-field theory for the bosons and Dynamical Mean-Field Theory for the fermions \cite{Titvinidze08}. 
The effect of the harmonic trap is treated within local density approximation (LDA): the harmonic potential is modelled by a spatially varying chemical potential. 
Exact diagonalization (ED) of the Anderson Hamiltonian is used as the DMFT impurity solver. 
To calculate the entropy we employ the thermodynamic relation $
d \Omega = - S dT - N_f d \mu_f - N_b d \mu_b  
$ for the grand potential $\Omega$.

\begin{figure}
\includegraphics[width=8.5cm]{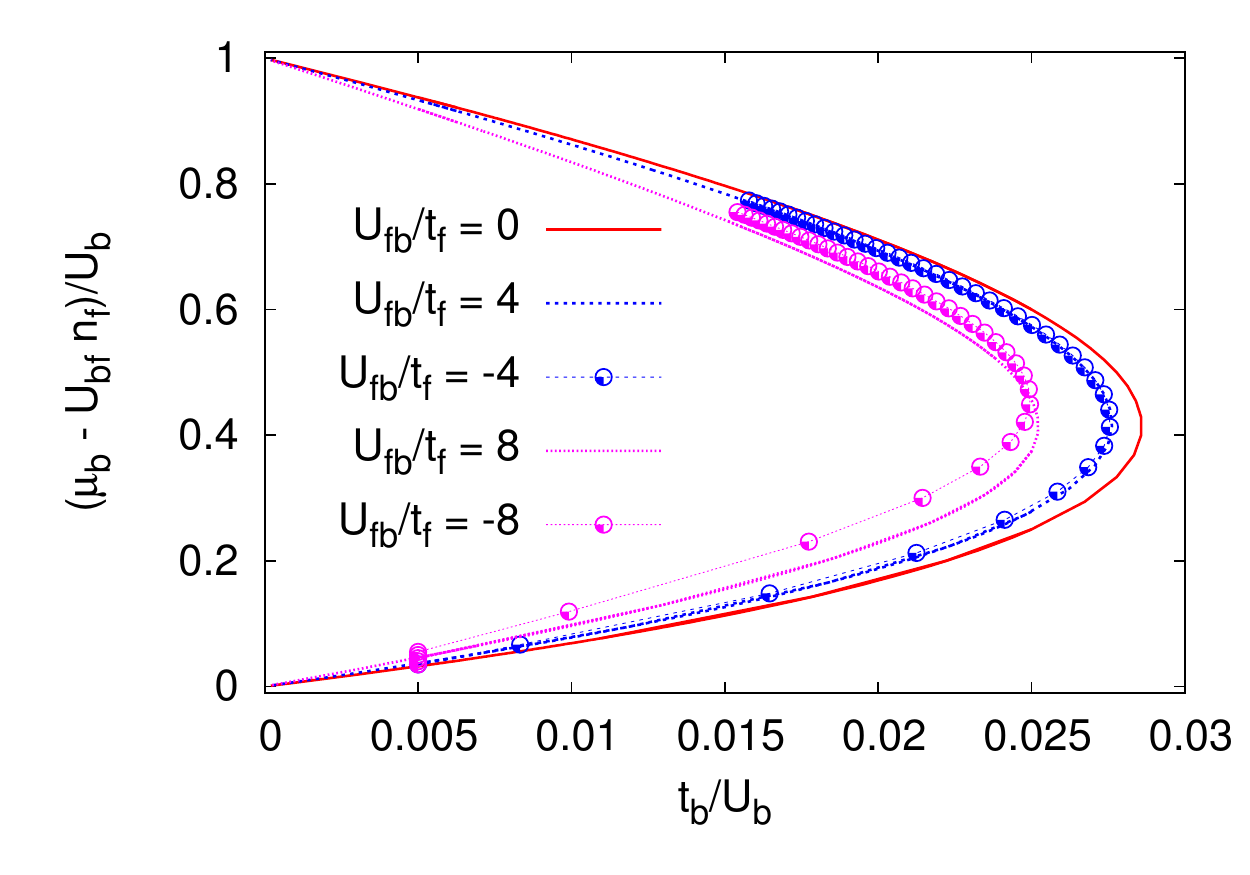}
\caption{(Color online) 
Phase border between 
bosonic Mott insulator and 
superfluid with $n_b=1$ 
for fermionic filling $n_f=0.7$, 
hopping ratio $t_f/t_b=2.2$, 
temperature $T \to 0$ 
and various values of $U_{bf}$.
The Hartree shift due to the Bose-Fermi interaction is subtracted from the bosonic chemical potential to facilitate an easy comparison.} 
\label{mottlobes}
\end{figure}

Before turning to the trapped mixture, we first present GDMFT results for a homogeneous system:
for constant fermionic filling $n_f=0.7$ the phase border between the superfluid and the Mott phase with bosonic filling $n_b=1$ is shown in Fig. \ref{mottlobes} for different values of the Bose-Fermi interaction. 
We observe that both for repulsive and attractive interaction the Mott lobe shrinks, thus giving evidence that the GDMFT scheme fully includes the dynamical screening of the bosonic interaction by the fermions.

\begin{figure}
\includegraphics[width=8.5cm]{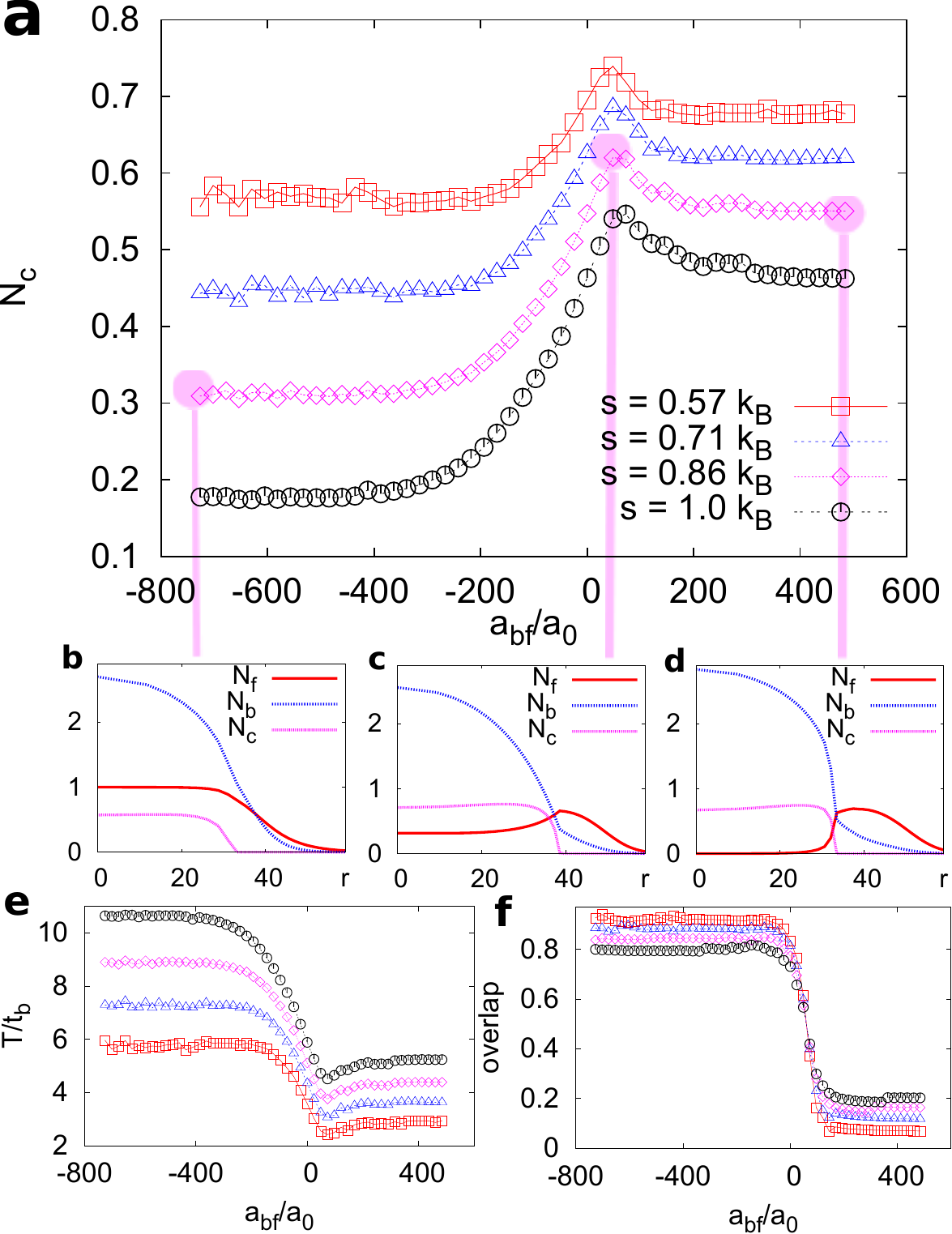}
\caption{(Color online) {\bf a} Total condensate fraction $N_c$, {\bf e} temperature $T$, and {\bf f} spatial overlap of the bosonic and fermionic clouds as a function of Bose-Fermi scattering length $a_{bf}$ for various values of the entropy per particle and $V_{\rm OL}^{\alpha}/E_R^\alpha = 9$ ($U_b/t_f = 5.2$, $V_f/t_f =0.0058$, $V_b/t_f = 0.0064$).
{\bf b}-{\bf d}: Spatial profiles of the bosonic and fermonic density and the local condensate fraction $N_c(r) = | \langle \hat b(r)\rangle |^2/N_b(r)$ as a function of distance from the trap center for three values of the Bose-Fermi interaction. 
 }
\label{s9}
\end{figure}

We now turn to the inhomogeneous case, for which we focus on the 
 condensate fraction of the bosons, which within LDA is obtained as 
\begin{equation}
N_c = \int d^3 {\bf r} |\langle b ({\bf r}) \rangle|^2 / \int d^3 {\bf r} n_b ({\bf r}).
\end{equation}
The experimentally measured visibility is a (monotonic) function of $N_c$ within GDMFT, therefore a maximum in the visibility corresponds to a peak in $N_c$.

The condensate fraction is calculated at constant entropy per particle $s = S_{\rm tot}/N_{\rm tot}$.
An estimate for the entropy per particle in the experiment is $s = 0.86 k_B$, but 
since this value is not known very accurately, we consider a range of values for $s$.

\begin{figure}
\includegraphics[width=4.65cm]{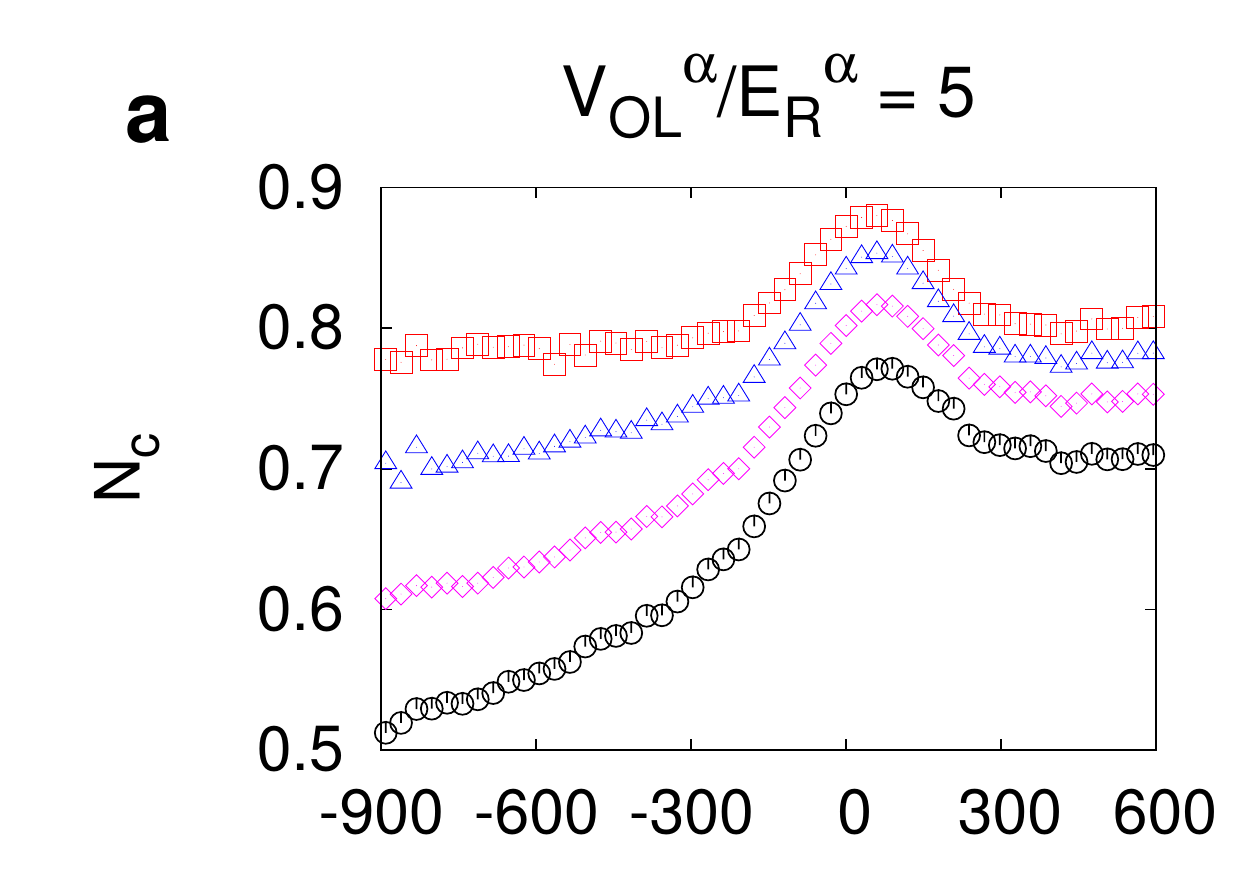} \hspace{-.9cm}
\includegraphics[width=4.65cm]{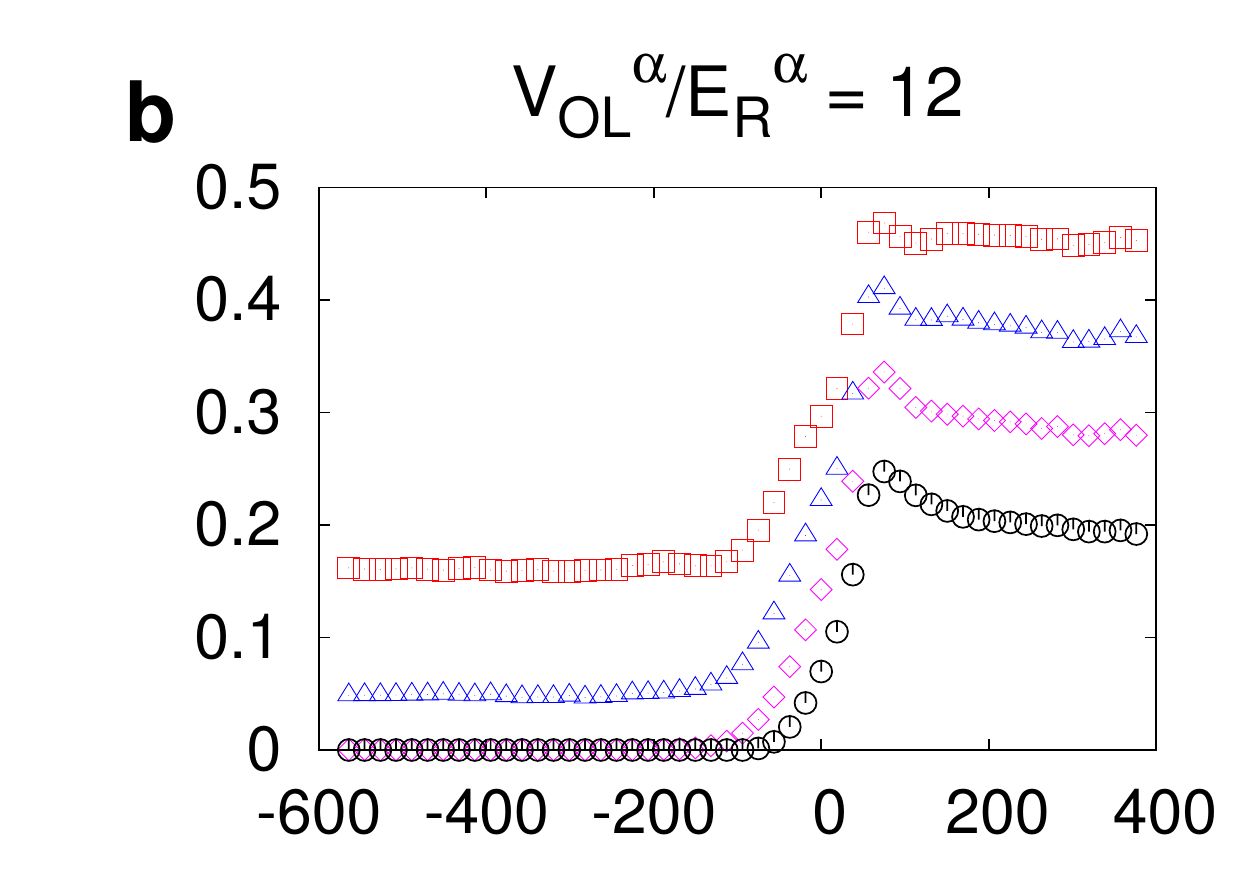}

\includegraphics[width=4.65cm]{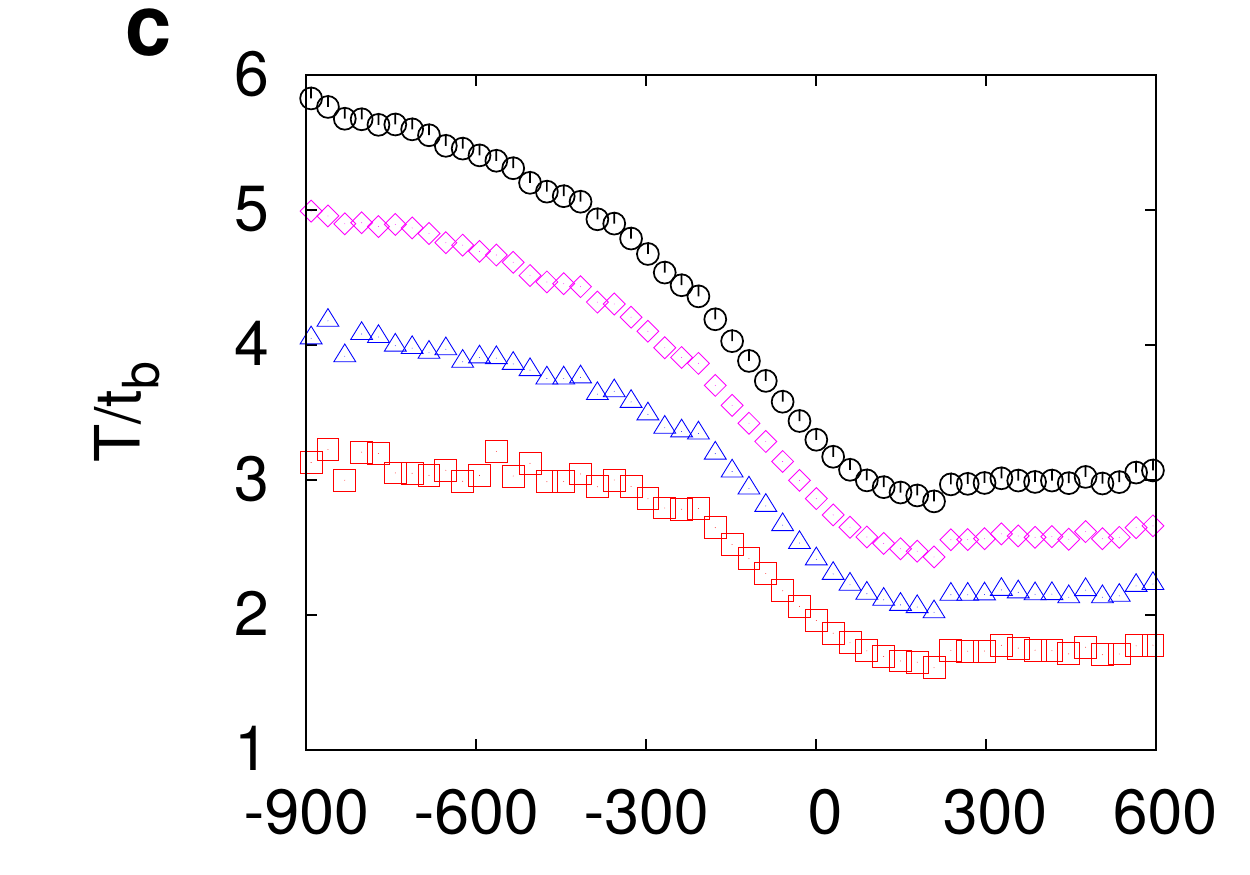} \hspace{-.9cm}
\includegraphics[width=4.65cm]{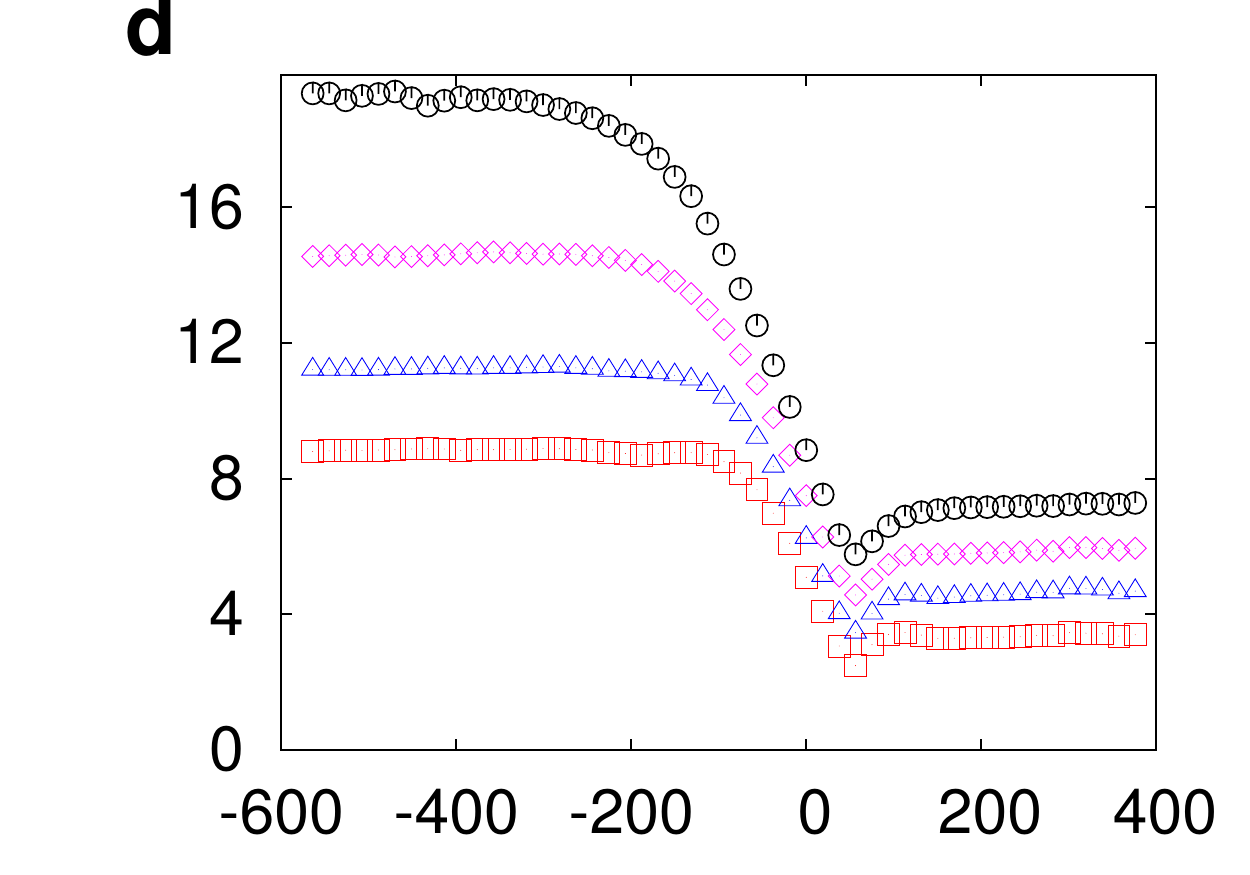}

\includegraphics[width=4.65cm]{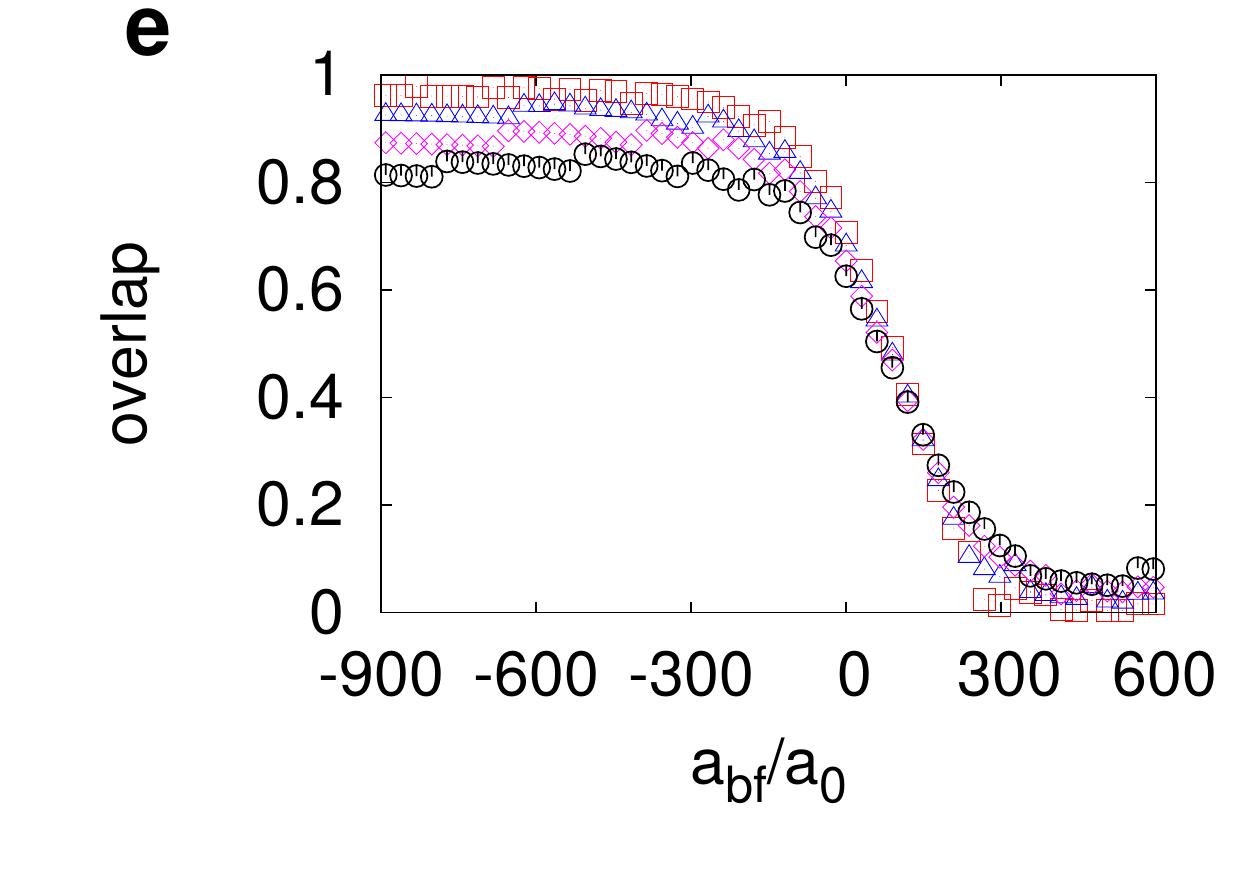} \hspace{-.9cm}
\includegraphics[width=4.65cm]{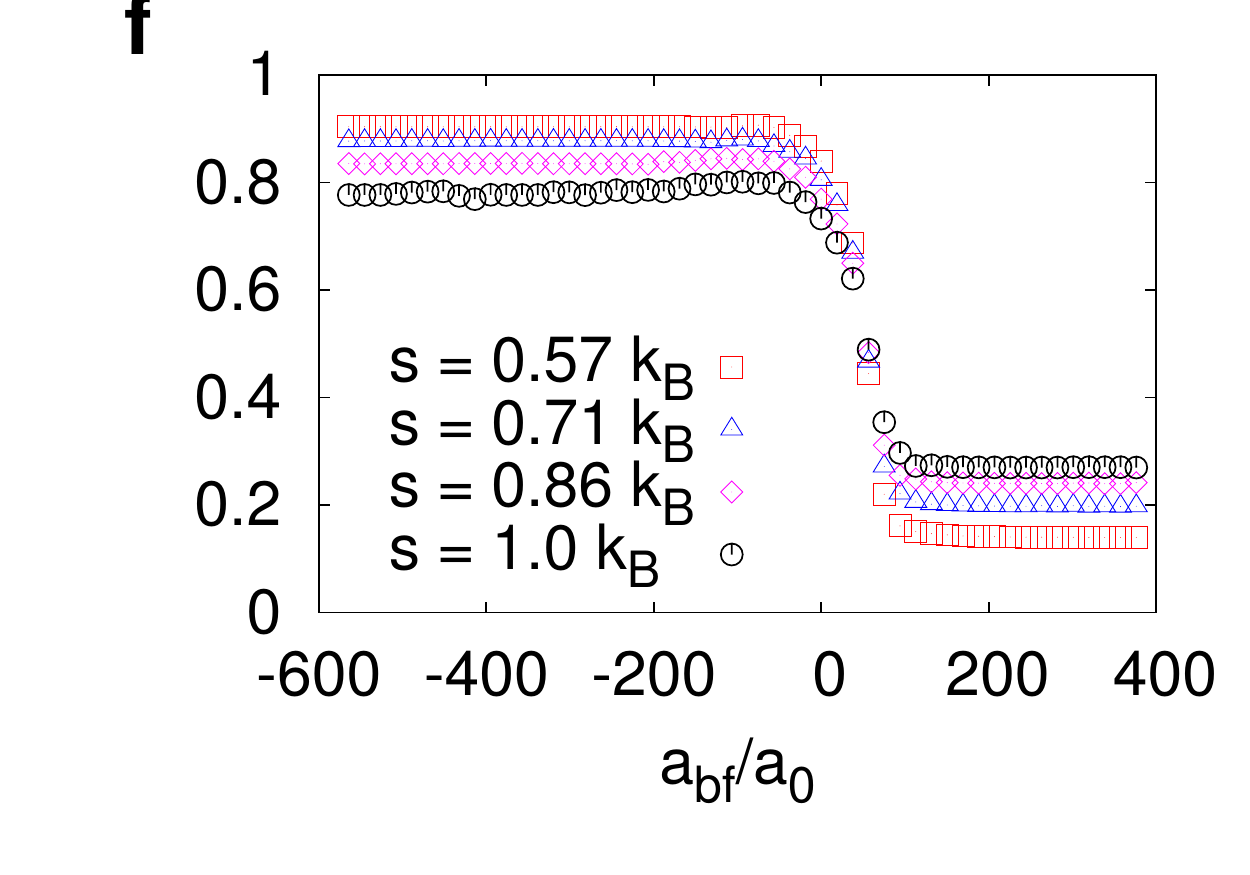}
\caption{(Color online) 
Condensate fraction (top row), temperature (middle row), and overlap of the bosonic and fermionic clouds (bottom row) for $V_{\rm OL}^{\alpha}/E_R^\alpha=5$ (left column) and $V_{\rm OL}^{\alpha}/E_R^\alpha=12$ (right column).  
Specific parameters are: $V_{\rm OL}^{\alpha}/E_R^\alpha=5$: $U_b/t_f = 1.06$, $V_f/t_f = V_b/t_f = 0.0024$;  $V_{\rm OL}^{\alpha}/E_R^\alpha=12$: $U_b/t_f = 13.4$, $V_f/t_f = 0.0108$, $V_b/t_f = 0.012$.}
\label{s5_12}
\end{figure}

We focus on the case of $V_{\rm OL}^{\alpha}/E_R^{\alpha} =9$ (shown in Fig. \ref{s9}), but data for $V_{\rm OL}^{\alpha}/E_R^\alpha=5$ and $V_{\rm OL}^{\alpha}/E_R^\alpha=12$ (see Fig. \ref{s5_12}) show similar features. Most importantly, the condensate fraction (Figs. \ref{s9}a, \ref{s5_12}a,b) shows a clear peak close to vanishing Bose-Fermi interaction for all lattice depths and for all values of the entropy considered, in agreement with the experimental observations \cite{Best_Bloch}. 

To get insight into the physical mechanism behind the peak in $N_c$, we plot the temperature at constant entropy in Fig. \ref{s9}e (cf. \ref{s5_12}c,d). Generally, the temperature has a minimum close to the position of the maximum in the condensate fraction $N_c$. 
In a qualitative way this minimum in the temperature at constant entropy can be understood as follows: for strong attraction only states in which bosons and fermions occupy the same lattice sites are energetically allowed.
This means that the size of the accessible Hilbert space is strongly reduced and hence the entropy at constant temperature is lowered. To obtain the same entropy, the temperature needs to be increased. The same mechanism works at strong repulsive interactions: in this case it is energetically unfavorable that a fermion and one or more bosons occupy the same site. This is an example of adiabatic heating if the Bose-Fermi interaction is changed. A similar effect has recently been observed in a fermionic mixture, where for increasing attractive interaction, the cloud expands due to a redistribution of entropy from spin to orbital degrees of freedom \cite{hackermueller10}.
In the Bose-Fermi mixture, the effect is larger on the attractive than on the repulsive side, leading to a higher condensate fraction for repulsive than for attractive interactions, in agreement with the experimental findings \cite{Best_Bloch}. This is because on the repulsive side the bosons and fermions phase separate, whereas for attractive interactions all the particles are packed in the center. This spatial redistribution is shown in the radial density distributions in Fig. \ref{s9}b-d where the density profiles for the bosonic and fermionic atoms and the local condensate fraction are plotted as function of the radius for strong attractive (b), weak repulsive (c), and strong repulsive (d) Bose-Fermi interaction. The profiles are plotted here at respective temperatures corresponding to total entropy equal to $s= 0.86 k_B$.
For strong Bose-Fermi attraction the bosons and fermions both occupy the center of the trap. The fermions form a band insulator; the bosons have a high filling and are superfluid. This implies that the bosons and fermions do not form bound states. If that were the case, the resulting composite particle would be fermionic in nature and the superfluid order parameter would vanish. 
For strong repulsion, on the other hand, the bosons occupy the center of the trap, whereas the fermions are pushed towards the outside. 

The most interesting regime is that of weak Bose-Fermi interaction, as shown in Fig. \ref{s9}c. The value of $a_{bf}$ chosen here corresponds to the peak in $N_c$. 
We observe that in this case the spatial distribution of the fermions is relatively flat.
This is due to the fact that for these parameters the interparticle repulsion partly compensates the harmonic confinement, which in turn leads to a large phase space 
and the possibility to store entropy efficiently 
as configurational entropy in the wings of the system.
We can carry this analysis one step further by observing that this value of $a_{bf}$ is approximately distinguishing the regime of phase separation of the bosonic and fermionic clouds (for stronger repulsion) from overlapping bosonic and fermionic clouds (for weaker repulsion and attraction). This is made clear in Fig. \ref{s9}f (cf. Fig. \ref{s5_12}e,f), where the spatial overlap of the bosonic and fermionic clouds (defined by the integral $ \int d^3 {\bf r} \; N_b ({\bf r}) N_f ({\bf r})$, here normalized to the total number of bosons $N_b$) is plotted at constant entropy. The sharp decrease in overlap approximately coincides with the minimum in temperature and the maximum in $N_c$. 
It is indeed reasonable to argue that at the onset of phase separation the phase space is largest, since then the bosons and fermions are neither bound to the same lattice sites, nor to excluding sites. This 
leads to a low temperature at fixed entropy, which in turn causes a high condensate fraction. 
We find the maximum in $N_c$, minimum in temperature, and onset of phase separation to happen at almost the same value of the interspecies interaction. However, the three do not exactly coincide, because of the complex interplay of many-body effects (leading also to a structure in $N_c$ at constant temperature when the interaction is changed) and the adiabatic heating effect explained above.  
 

In conclusion,
we have investigated a trapped Bose-Fermi mixture by means of GDMFT. 
The dynamical screening of the Bose-Bose interaction by the fermions leads to decreasing Mott insulating lobes in the ground state of homogeneous mixtures. 
In the trapped case, a peak in the condensate fraction is obtained close to vanishing Bose-Fermi interaction. 
We have shown that the dramatic decrease in condensate fraction for attractive Bose-Fermi interactions is caused by adiabatic heating due to the strong reduction of the available Hilbert space. For both positive and negative interspecies interaction strength we find a strong redistribution in the density of the two-component gas. 
The combination of both effects leads to the peak in the condensate fraction close to vanishing Bose-Fermi interactions.
We remark that these effects are universal and important for the interpretation of a large class of experiments 
with ultracold atoms in optical lattices.

Although we thus obtained good qualitative agreement with the experimental results, there is no full quantitative agreement. In particular, 
it was observed that the peak in the visibility occurs exactly at vanishing Bose-Fermi interaction, whereas we find it to happen at a small, but positive value of the interaction. There are a number of possible explanations for this disagreement.
First of all, our analysis relies on the assumption of adiabatic loading of the optical lattice. 
Even for moderate interactions, however, one can expect mass transport to be affected, leading to longer adiabaticity timescales, as evidenced in recent transport experiments \cite{Schneider10, Hung10}.  
The strong particle redistribution for large attractive and repulsive interaction can easily induce additional heating and shift the peak in the condensate fraction. 
A second effect not included in our study is the presence of gravitational sag between the bosonic and fermionic cloud in the experiment, due to their different masses. 
This separates the centers of mass of the two clouds already without interspecies interactions 
and additional repulsion enhances this separation.
Since we observe in our calculations that the point where the clouds separate is correlated with the maximum in the condensate fraction, the experimental sag is likely to shift the peak in the visibility towards smaller Bose-Fermi interaction.
Finally, residual uncertainties in the scattering length or multi-band effects could play a role in the exact location of the peak in the condensate fraction. Although large qualitative effects are only expected for strong interspecies interactions \cite{Luhmann08, Lutchyn08, Tewari09, Mering10}, at small values of $a_{bf}$ multiband effects can still lead to renormalization of the parameters and small quantitative changes in the predictions. 

We thank Sebastian Will and Thorsten Best for useful discussions.
This work was supported by the Netherlands Organization for Scientific Research (NWO), the German Science Foundation DFG (FOR 801, 
SFB/TR 49 and the DIP project BL 574/10-1)
and the National Science Foundation 
under Grant No. NSF PHY05-51164.

\emph{Note added.--} During completion of this work, related results were reported in \cite{Cramer10}.

\end{document}